# Revealing Phonon Bridge Effect for Amorphous vs Crystalline Metal–Silicide Layers at Si/Ti Interfaces by a Machine Learning Potential

Mayur Singh*[1], Lokanath Patra[1], Chengyang Zhang[2], Greg MacDougall[3], Suman Datta[2], David Cahill[3] and Satish Kumar†[1]

[1]*George W. Woodruff School of Mechanical Engineering, Georgia Institute of Technology, Atlanta, GA;*

[2]*School of Electrical and Computer Engineering, Georgia Institute of Technology, Atlanta, GA;*

[3]*Department of Materials Science and Engineering, University of Illinois Urbana-Champaign, IL*

**ABSTRACT**. Metal–semiconductor interfaces play a central role in micro- and nano-electronic devices as heat dissipation or temperature drop across these interfaces can significantly affect device performance. Prediction of accurate thermal boundary resistance (TBR) across these interfaces, considering realistic structures and their correlation with underlying thermal transport, remains challenging. In this work we develop a unified Neuroevolution Potential (NEP) for the Si–Ti system that accurately reproduces energies, forces, and phonon properties of bulk Si, Ti, and $TiSi_2$ and extends naturally to interfacial environments to analyze interfacial transport. An important development over current machine-learned interatomic potentials is the capability to model complex structures at metal-semiconductor interfaces, as the NEP enables large-scale non-equilibrium molecular dynamics simulations of epitaxial Si/Ti interfaces to elucidate the effect of amorphous or crystalline silicide interfacial layers. Simulated TBRs show excellent agreement with our time-domain thermoreflectance (TDTR) measurements, validating the robustness of our predictions. Spectral analyses reveal that amorphous $TiSi_2$ interfacial layer helps in efficient interfacial transport when the thickness is less than 1.5 nm compared to the crystalline $TiSi_2$ layer, due to the high spectral conductance in the 3-6 THz frequency range and also due to the opening of channels for anharmonic transport, but this trend reverses when the interfacial layer thickness increases beyond 1.5 nm. Comparison of TBRs at Si/$TiSi_2$ interface for different crystalline phases of $TiSi_2$ establishes that C54 phase has reduced TBR compared to C49 phase, which is correlated with the difference in their phonon density of states (PDOS) overlap with Si. These results provide atomistic insight into the role of crystalline versus amorphous silicides in interfacial heat transport and demonstrate a transferable machine-learned potential for studying heat dissipation in advanced semiconductor devices.

## I. INTRODUCTION.

Understanding heat transfer across metal–semiconductor interfaces remains a critical challenge for the design of nanoscale electronic devices. The energy transport mechanism across such interfaces and its dependence on interfacial structures are still not fully understood. Methods to quantify resistance at the interface as the interfacial thermal conductance or its inverse thermal boundary resistance (TBR), known as Kapitza resistance, have long been developed [1-3]. TBR at interfaces is governed specifically by the scattering of phonons with the constituent materials of the interface. Several approaches have been developed to model thermal transport across metal–semiconductor interfaces. Theoretical models such as the acoustic mismatch method (AMM) [4], and diffuse mismatch method (DMM) [5], have been used for TBR predictions at interfaces, typically considering only elastic phonon transport and neglecting the atomistic structure of the interface. Methods to include inelastic scattering have been developed [6] but do not lead to sufficient predictive tools for analyzing interfacial heat transfer accurately. The Atomistic Green's function (AGF) is one of the methods that considers the atomistic structure of the interface, and has enabled the modeling of a wide range of materials [7-11], considering inputs of interatomic force constants from the first-principles density functional theory (DFT) or from the empirical potentials. These studies also often neglect the disorder at the metal–semiconductor interface and rely on idealized, atomically sharp geometries. Majumdar et al. [12] established that the TBR at the metal-semiconductor includes can be significantly increased by weak electron-phonon coupling, but the significance of such coupling has not been established across all metal-semiconductor systems. Formulations and corrections have been developed for considering electron-phonon coupling and/or anharmonicity in the AGF simulations

*Contact author: msingh96@gatech.edu

†Contact author: satish.kumar@me.gatech.edu

[8, 13], but applications of these AGF models are limited to ideal interfaces, and anharmonic scattering is either neglected or limited to approximate methods such as Buttiker probe scattering [14]. Experimental measurements using TDTR comparing epitaxial and nonepitaxial silicide/silicon interfaces suggest that TBR is not very different, and there is no significant effect of different interface orientations for epitaxial interfaces, while AGF significantly overpredicts TBR above 100K [15]. Additionally, it remains open-ended how significantly inelastic phonon scattering or electron-coupling affect transport mechanisms at metal–semiconductor interfaces with disorder [14].

Thermal transport in systems with more complex interfaces can be modelled using Molecular Dynamics (MD) simulations [16], which considers anharmonic scattering by default. Metal-semiconductor interfaces have been extensively studied using nonequilibrium molecular dynamics (NEMD) simulations [17-22]. Methods to decouple and evaluate relevant scattering mechanisms in MD have been explored by several groups [23-25]. However, significant discrepancies between simulations and experiments have been noted due to the choice of interatomic potential used to perform the MD simulations. Empirical potentials like the modified embedded atom method (MEAM) potential [26], have underestimated the TBR for Al/Si [22] interfaces. While manually fitted parameters can sometimes be adjusted to improve agreement with measurements, doing so typically compromises the ability to accurately capture the phonon properties of the bulk materials and the metal–semiconductor interface itself.

Machine-learned interatomic potentials (MLIPs) have emerged as powerful tools for more accurate thermal simulations using MD. First principles DFT data can be used to train models like neural networks to predict the interatomic forces and potential energy surfaces based on the local atomic structure [27]. The training of interatomic force fields in this manner has been shown to obtain accurate phonon properties, including more accurate lattice thermal conductivity and TBR [28-32]. Several approaches for MLIPs have been developed, including the Voxelized Atomic Structure (VASt) [32, 33] potential, the Deep potential for molecular dynamics (DeePMD) [34], and the Neuroevolution Potential (NEP) [35]. Notably, a DEEP potential was used to study the interfacial heat transfer between the Al/Si interface by Khrot, K., et al. [18], which marks a significant promise in accurate modelling of metal/semiconductor interfaces using MLIPs. Though more complex interfaces can present a larger variety of atomic systems to be considered, including crystalline and amorphous phases. MLIPs have not been developed for metal silicides and their interfaces. The difference in transport mechanism between thin films of crystalline vs amorphous phases of the silicide at the interface is not understood.

The Si/Ti interface is prominently used in VLSI technology [36] as a metal contact for integrated circuits and field-effect transistors (FETs), because of its ability to form silicide. Several crystalline and amorphous phases of the silicide exists, e.g., three crystalline $TiSi_2$ phases, including C49, C54, and C40 phases [37, 38], though the C49 and C54 are more commonly identified [39]. The interfacial thermal transport at the Si/TiSi2-C49 and Si/TiSi2-C54 interfaces has been investigated using AGF and experimental methods [14, 15, 40]. Modelling of interfaces that include the disordered or mixed phases at the Si/Ti interface, as is typical in experimental systems, remain unexplored. Due to the large feature space of interface structure that needs to be considered, developing an MLIP of the Si/Ti systems requires a large dataset to accurately explore the thermal transport.

In this work, we develop a Neuroevolution Potential (NEP) to accurately capture the energetics and atomic forces of Si/Ti systems, enabling large-scale MD simulations of interfacial heat transport. Using this potential, we compute the thermal boundary resistance (TBR) for a range of Si/Ti interfaces, including epitaxial, amorphous, and silicide-forming configurations. We fabricate samples of Si/Ti with thin amorphous layer at the interface and perform TBR measurements using time-domain thermo-reflectance (TDTR). The excellent agreement against experimental measurements validates the robustness of predictions from MD simulations using developed MLIPs. We find that considering only phonon-phonon scattering brings good agreement with experiments for Si/a-TiSi/Ti systems, showing weaker than previously suggested electron phonon-coupling in Si/metal silicide interfaces [40]. We analyze spectral phonon contributions to heat transfer and establish that a ~0.5 nm thin amorphous TiSi structure at Si/Ti interface has reduced TBR compared to a crystalline $TiSi_2$ structure. We evaluate the impact of interface thickness and observe that the trend of TBR reverses between crystalline vs amorphous insertions at Si/Ti interface once the insertion thickness is higher than 1.5 nm. We also compare TBRs at $Si/TiSi_2$ for different crystalline phases of $TiSi_2$ and observe that C54 phase has reduced TBR compared to C49 phase, which can be understood by the difference in their phonon density of states (PDOS) overlap with Si and spectral heat fluxes.

*Contact author: msingh96@gatech.edu

†Contact author: satish.kumar@me.gatech.edu

## II. METHODS

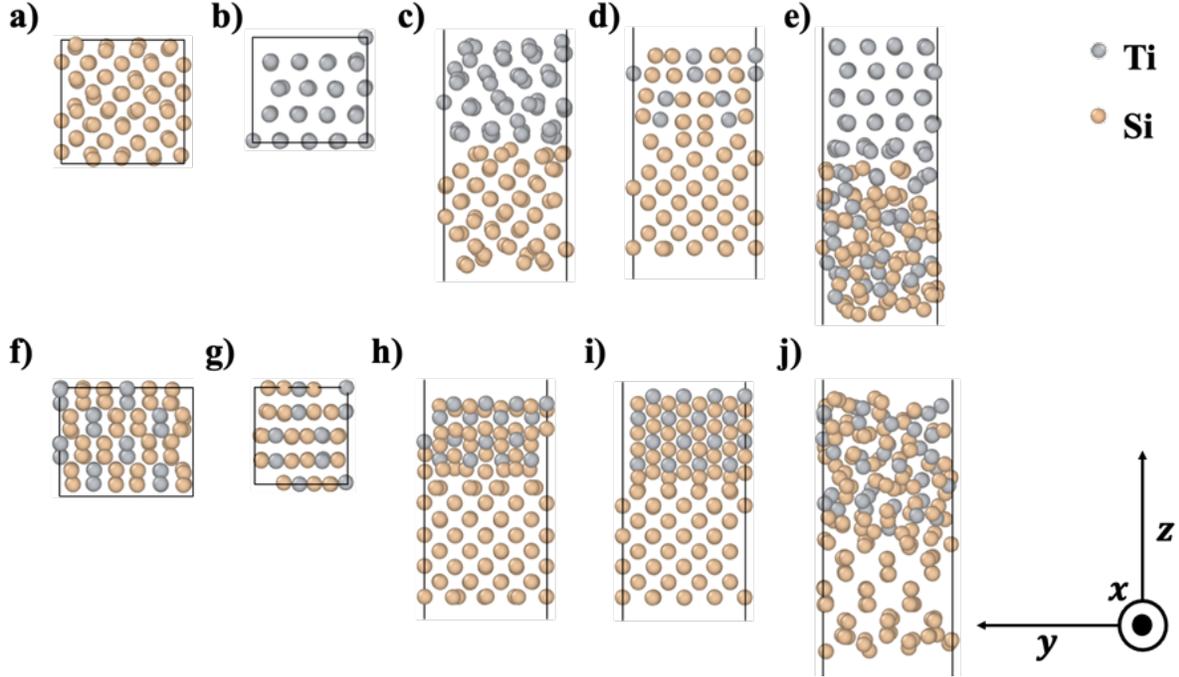

FIG. 1. Snapshots of some structures included in the training data of the NEP.

### A. Training Dataset

Density functional theory (DFT) calculations were carried out using the VASP package to generate total energies, atomic forces, and virial stresses for a diverse set of structures, including elemental Si and Ti, the C49 and C54 phases of TiSi$_2$, amorphous TiSi$_x$ (a-TiSi$_x$), and representative Si/Ti interface structures, as shown in Fig. 1. The training dataset was constructed using ab initio molecular dynamics (AIMD) simulations across a variety of systems relevant to the Si/Ti interface. The main phases included bulk silicon supercell, a bulk Ti [(0)001] surface, and 6 different surfaces of crystalline silicide (three each for the C49 and C54 phases). Atomic configurations for interfacial structures included combinations of Si/Ti, Si/TiSi$_2$, Si/a-TiSi$_x$, and a-TiSi$_x$/Ti. The titanium surface was stretched to 4.7% to match the silicon lattice constant, and the silicide surfaces were stretched to 2.5% to match silicon. Stretching was not needed for binary interfaces with amorphous solids. Additional details of the dataset are provided in Table 1.

To sample atomic configurations, AIMD simulations were performed in the isothermal-isobaric ensemble(NPT ) at 300 K for 5 ps to capture changes in structure due to thermally relevant atomic vibrations. For the samples with interfaces, both room-temperature and melt-quench simulations up to 1500 K were conducted to sample possible disorder and interdiffusion effects that may arise during processing or operation. These simulations were 3 ps in length, with atomic configurations sampled every 10 steps to ensure decorrelated structural snapshots. In total, the dataset comprises 4016 atomic configurations spanning crystalline, amorphous, and interfacial environments, including both relaxed and thermally perturbed structures.

TABLE I. Details of the training data used to develop NEP.

| System | No. of atoms | structures |
|---|---|---|
| Bulk Si | 8, 64 | 170 |
| Bulk Ti | 32,48,64 | 200 |
| Si/Ti interfaces | 136 | 1436 |
| TiSi$_2$-C54 surfaces | 48,96 | 450 |
| Si/TiSi$_2$-C54 interfaces | 112,160 | 450 |
| TiSi$_2$-C49 surfaces | 108 | 30 |
| Si/TiSi$_2$-C49 interfaces | 112,160 | 900 |
| Ti/a-TiSi interface | 128 | 280 |
| Si/a-TiSi interface | 180 | 100 |

### B. DFT Calculations

*Contact author: msingh96@gatech.edu

†Contact author: satish.kumar@me.gatech.edu

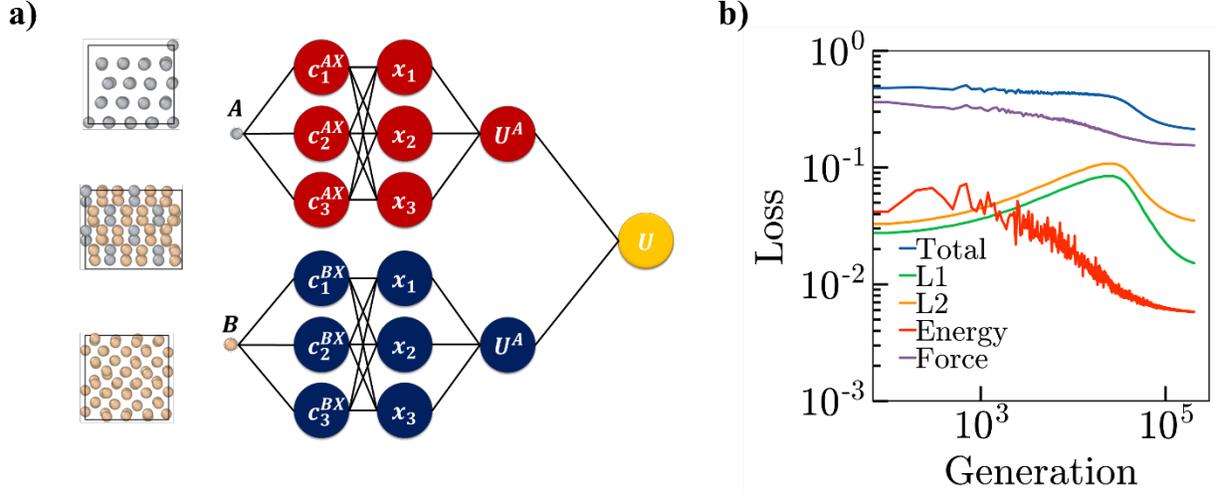

FIG 2. Illustration of NEP4 architecture for a 2-component system. A and B are the atom types, $c_x$ are descriptors, x represents activation functions, and U is the atomic energy; and b) Losses of the training dataset over $4e^5$ generations. L1 and L2 are the regularization functions of the loss.

Single-point calculations were calculated at the Perdew–Burke–Ernzerhof (PBE) exchange-correlation level. A plane-wave basis with a cutoff energy of 600 eV was chosen to ensure reliable convergence. For Brillouin zone integration, a Γ-centered k-point mesh of 4×4×4 was applied to bulk configurations, while a coarser 2×2×1 mesh was used for interfacial supercells to account for reduced periodicity and increased system size. All calculations were converged to at least 10-8 eV in the electronic self-consistency loop to guarantee precise evaluation of interatomic forces and stress tensors.

### C. NEP Model

Using the generated training data, we train a Neuroevolution potential (NEP) as implemented in the Graphics Processing Unit Molecular Dynamics (GPUMD) code. NEP is a Neural Network Potential (NNP), a class of MLIPs which use feed-forward artificial neural networks (ANN) as the regression kernel. The ANN represents site energy $U_i$ of atom $i$ as a function of a descriptor vector with $N_{des}$ components:

$$U_i = \sum_{\mu=1}^{N_{neu}} \omega_\mu^{(1)} \tanh\left(\sum_{\nu=1}^{N_{des}} \omega_{\mu\nu}^{(0)} c_\nu^i - b_\mu^{(0)}\right) - b^{(1)} \quad (1)$$

where $N_{neu}$ is the number of neurons in the hidden layer, $\tanh(x)$ is the activation function, $\omega^x, b^x$ are the weights and biases, and $c_\nu^i$ represents the descriptors, which include separately derived radial and angular many-body terms. Fig. 2(a) shows a general representation of the NN architecture, where $c_i^{AX}$ represents the descriptors of atom $A$ paired to atom type X (in this case A or B), and $x_j$ are the activation functions shown in Eq. (1). This addition is unique to the NEP version 4 (NEP4) [41] framework. NEP4 was initially introduced for modeling high-entropy alloys, but here we propose its application to complex Si/Ti interfaces. These interfaces often exhibit significant structural roughness, due to the self-aligned silicide process, where the deposited metal on semiconductor material of the channel, to form source or drain contacts, reacts to form metal silicide films in the device [42]. The Separable Natural Evolution Strategy (SNES) [43] employed in the NEP framework facilitates exploration of the configuration feature space well beyond the initial training dataset. The optimization of NEP model parameters is achieved by minimizing a loss function that includes a weighted sum of root mean square errors (RMSEs) for energy, forces, and virials, alongside regularization terms that prevent overfitting. Regularization, implemented through both L1 and L2 penalties. Figure 2(b) illustrates the evolution of the total loss during training, as well as the individual contributions from the L1 and L2 regularization terms, together with the RMSEs for potential energy, interatomic forces, and atomic virials on the training dataset.

### D. NEMD simulations

Nonequilibrium molecular dynamics (NEMD) simulations were performed using GPUMD. The system was first equilibrated using the NPT ensemble at zero pressure in the x and y directions, modeling an

*Contact author: msingh96@gatech.edu

†Contact author: satish.kumar@me.gatech.edu

interface between bulk Si and the Ti [0001] surface. After at least 1 ns of equilibration, NVT thermostats are applied to two regions, each 4 nm in length, at the simulation box boundaries to induce a heat flux, as illustrated in Fig. 3(a). To estimate the thermal

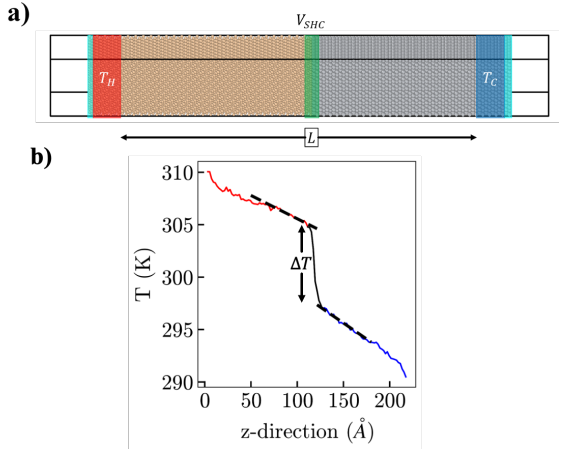

FIG 3. a) Representative schematic of NEMD simulations. $V_{SHC}$ is the interfacial region that is used to estimate interface resistance. $T_H$ and $T_C$ are temperatures of thermostat regions; and b) a representative temperature profile is used to estimate temperature drop $\Delta T$ and TBR at the interface

boundary resistance (TBR), the resulting temperature drop $\Delta T$ across the interface was measured, as shown in Fig. 3(b). The thermostat temperatures $T_H$ and $T_C$, are 310 and 290 K, respectively. The production run for these simulations was 10 ns, and 5 independent simulations were performed to quantify the uncertainty of the TBR measurements. The TBR is then computed as the inverse of the interfacial thermal conductance:

$$R = \frac{\Delta T}{Q/A} \qquad (2)$$

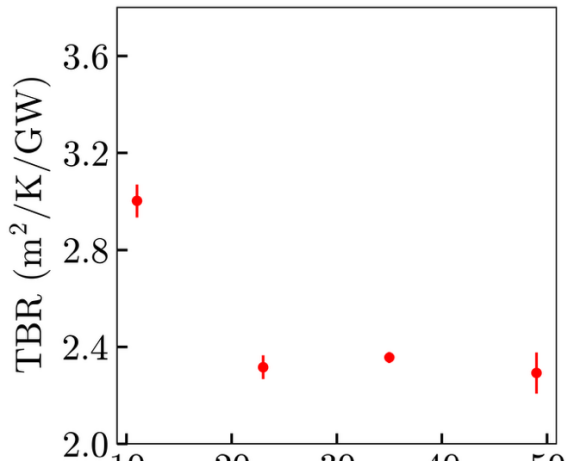

Figure 4. TBR of Si/Ti interface with increasing sample length L.

*Contact author: msingh96@gatech.edu

†Contact author: satish.kumar@me.gatech.edu

Where $Q$ is estimated from $Q = \frac{dE}{dt}$, which is the rate at which energy is added to and subtracted from the hot and cold baths.

NEMD simulations are known to exhibit size effects, particularly with respect to the sample length $L$ between hot and cold regions. The effect of the lateral size area was negligible past cross-sectional areas of ~9 nm², a cross sectional dimension of ~4.3x4.3 nm was chosen to minimize strains from lattice mismatch. To determine a suitable length scale for our study, we performed NEMD simulations on the Si/Ti epitaxial interface with $L$ varying from ~10 nm to ~50 nm. We observed that the thermal boundary resistance (TBR) decreased with increasing length and converged to approximately $2.3 \, m^2 K \cdot GW^{-1}$ beyond $L$=20 nm, as shown in Fig. 4. Based on this, 20 nm was selected as the sample length for all subsequent NEMD simulations to ensure reliable and size-independent results.

### E. Spectral Conductance of NEMD simulations

To further analyze transport mechanisms, we can decompose the thermal transport at the interface using the spectral decomposition of the heat current proposed by Sääskilahti, K. et al [45], where we monitor the viral velocity correlations of the interfacial atoms, which are defined by: potential stress and velocity correlation function $K(t)$: $K(t) = \sum_i \langle W_i(0) \cdot v_i(t) \rangle$. By taking the Fourier transform we obtain:

$$\widetilde{K}(\omega) = \int_{-\infty}^{\infty} dt e^{i\omega t} K(t) \qquad (3)$$

And setting $t = 0$ we can obtain the spectrally decomposed heat current:

$$J(\omega) = \int_0^{\infty} \frac{d\omega}{2\pi} [2\widetilde{K}(\omega)] \qquad (4)$$

From which we can derive a spectrally decomposed TBR:

$$R = \frac{\Delta T}{J(\omega)/V} \qquad (5)$$

We decompose the conductance of the interfacial region, as shown in Fig. 3(a), and determine which phonon modes from either side of the interface contribute to the thermal transport.

### F. Sample Fabrication and Time Domain Thermoreflectance Measurements

To validate the interfacial NEP model, simulation results for thin a-TiSi layer at the interface were compared against time-domain thermoreflectance (TDTR) measurements on Si/Ti/Al samples,

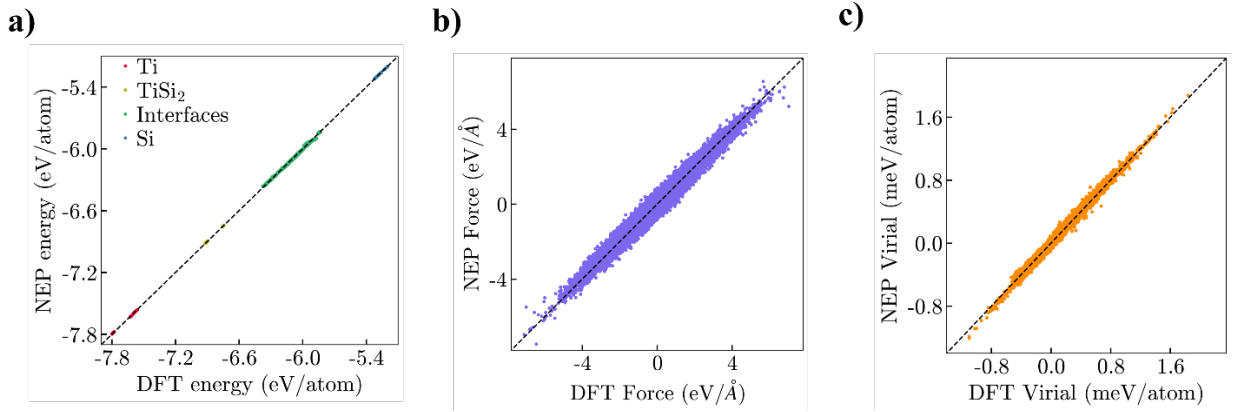

Figure 5 a) Energy RMSE of the NEP, clustered to represent the different systems in the training data; b) Force RMS; and c) Virial RMSE.

performed at UIUC These samples were prepared on low-doped n-type Si substrates (~1×10$^{16}$ cm$^{-3}$) to minimize electronic contributions to the TBR. After RCA surface cleaning, 15 nm Ti and ~72–80 nm Al layers were deposited using sputtering under high vacuum conditions. Thermal annealing at 450 °C for 1–30 minutes in N$_2$ atmosphere produced TiSi$_x$ interlayers with varying thickness and crystallinity. Tunneling Electron Microscopy (TEM) and energy dispersive X-ray spectroscopy (EDS) were performed to see the thickness of the resulting silicide layer present in the interface. The thermal boundary resistance (TBR) of Al/Ti-silicide/Si interfaces was measured by time-domain thermoreflectance (TDTR) at the U. of Illinois. TDTR employs a high repetition rate (76.8 MHz) mode-lock Ti:sapphire laser to generate a modulated pump beam (9.3 MHz) and a time-delayed probe beam that heat the surface of the sample and detect transient changes in the surface temperature. A 5× microscopy objective was used to focus the pump and probe beams to a 1/e$^2$ intensity radius of 10 μm. Sharp-edged optical filters were used to spectrally separate the pump and probe beams. The average powers of the pump and probe beams were 14 and 6 mW, respectively. TDTR data in the form of the negative of the ratio of the in-phase and out-of-phase signals detected synchronously with the pump modulation frequency were fit by a numerical solution of the heat diffusion equation of a multilayer sample, where each layer is modeled by a thermal conductivity, volumetric heat capacity, and layer thickness. The only free parameter is the interface layer that is modeled as a thin layer (1 nm) with a small volumetric heat capacity (0.1 J/(cm$^3$ K)) and unknown thermal conductivity. The atomic areal densities of Al and Ti were measured by Rutherford backscattering spectrometry, and the metal layer was modeled as a single layer using a weighted average of the heat capacities of Al and Ti.

## III. RESULTS

### A. Model Validation

As shown in Fig. 5. (a-c), the trained NEP has a root mean square error (RMSE) of 5.9 meV/atom for energy, 159 meV/Å for forces, and 30 meV/atom for virial. Because our dataset includes rough/disordered interfaces, we obtain an energy RMSE that is higher than typically reported in literature while using NEP potential, but reasonably accurate for studying these interfaces together. We additionally evaluate the efficacy of the NEP by calculating the phonon dispersions of bulk Si and hcp-Ti and comparing the results with phonon dispersions obtained from the DFT simulations or experiments.

For bulk silicon, the harmonic force constants for a 5×5×5 silicon supercell were obtained using the NEP, followed by computation of the phonon dispersion, as shown in Fig. 6(a). The acoustic branches exhibit excellent agreement with both DFT and experimental data, particularly near the Γ point. The optical phonons show a modest underprediction, with the highest optical branch approaching ~15 THz, slightly below the 15.4 THz predicted by DFT and observed experimentally [44]; similar trends have also been reported for previously reported NEPs [45]. Also of note is the larger gap predicted by NEPs between optical and acoustic branches at the L point in the Brillouin zone. The excellent matches between phonon dispersion predictions by DFT and NEP, as shown in Fig. 6, increase confidence in the predictive capability of the developed NEP.

*Contact author: msingh96@gatech.edu

†Contact author: satish.kumar@me.gatech.edu

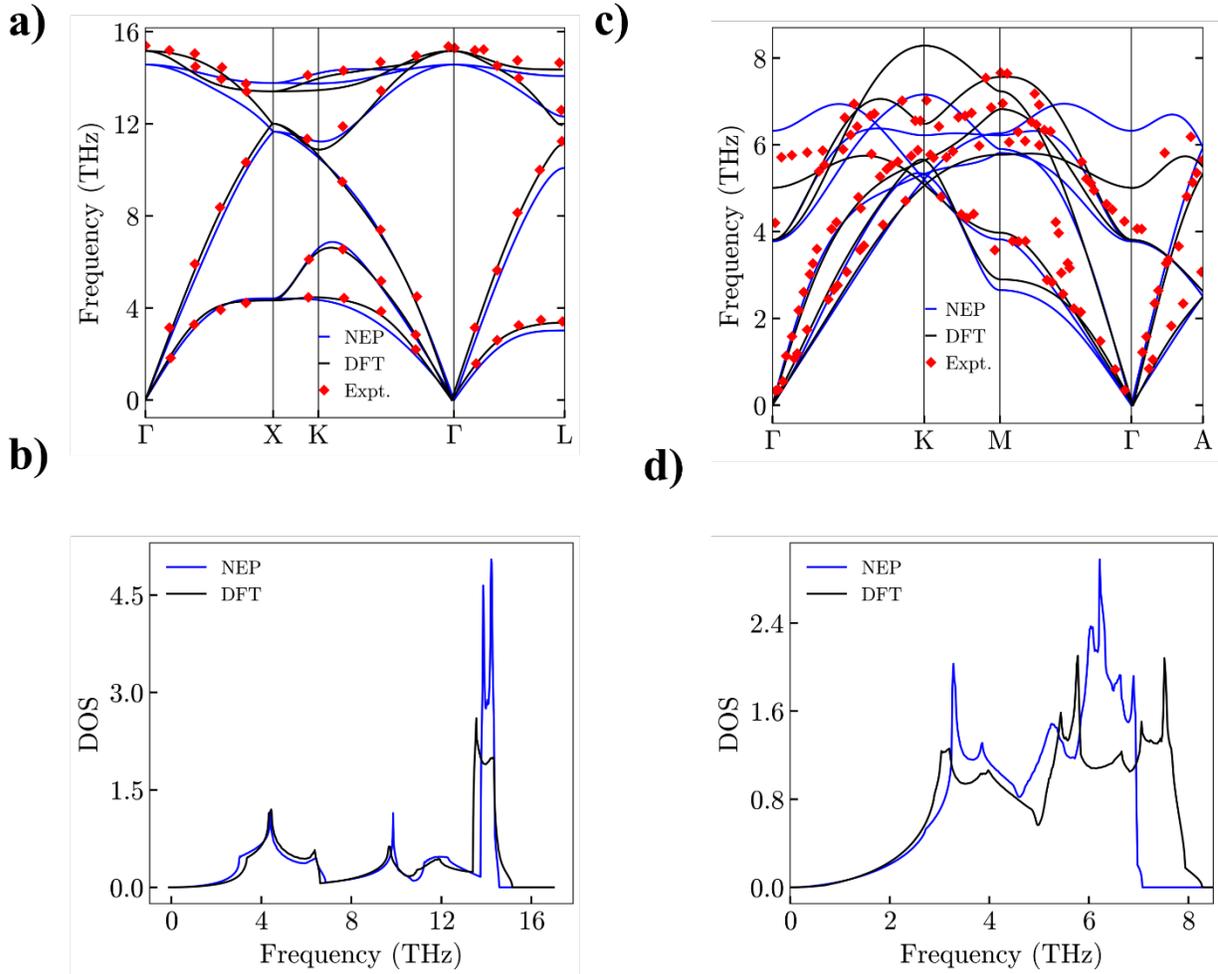

Figure 6 Comparison of NEP predicted dispersion curves and PDOS with DFT simulations or experimental measurements. a) Phonon dispersion of bulk Si; b) Phonon Dispersion of hcp Ti; c) PDOS of Si; and d) PDOS of Ti.

The phonon dispersion of hcp-Ti (α-phase), as shown in Fig. 6(b), agrees well with experimental data at low frequencies, particularly along the Γ–K direction corresponding to the [0001] surface. NEP predictions agree well with the overall trend of the experimental data [46], especially for acoustic phonons, but they overestimate the frequencies of optical phonons at the Γ point and underestimate the frequencies of optical phonons at the K and M points. This is consistent with prior NEPs produced for Ti systems [47] as well as other MLIPs [48]. In the context of comparing the results from different methods, the disagreement between experiment and DFT has already been noted in the literature [49]. Considering even the first principle DFT prediction does not match with experiments exactly for the two pure elements, we proceed with using developed NEPs for analyzing TiSi$_2$ phonon properties and thermal transport investigation by the NEMD studies.

Lastly, we obtained the phonon dispersions of both C54 and C49 TiSi$_2$. To date, only the phonon properties of C49 TiSi$_2$ have been reported, most notably by Sadasivam et al [40] and no experimental data on the phonon dispersion of TiSi$_2$ exists, particularly for the C54 phase, which we report in this work. Orthorhombic K-paths were used to plot the PDOSs, and additional DFT calculations were performed for comparison. Overall, excellent agreement between DFT and NEP predictions is observed. For the C54 TiSi$_2$'s PDOS and phonon dispersion of the C54 phase of TiSi$_2$, we observe the emergence of a small phonon bandgap from NEP predictions. While this bandgap disappears under NEMD simulations, its prediction in the lower-symmetry C54 phase TiSi$_2$ has not been reported before. The literature suggests that a phonon bandgap if significant, suppresses scattering between acoustic and optical modes [50], thereby enhancing thermal conductivity. Such gaps are often attributed to mass differences within the lattice [51], and their appearance in C54 TiSi$_2$ underscores the potential of silicide phases to be engineered for improved transport properties.

*Contact author: msingh96@gatech.edu

†Contact author: satish.kumar@me.gatech.edu

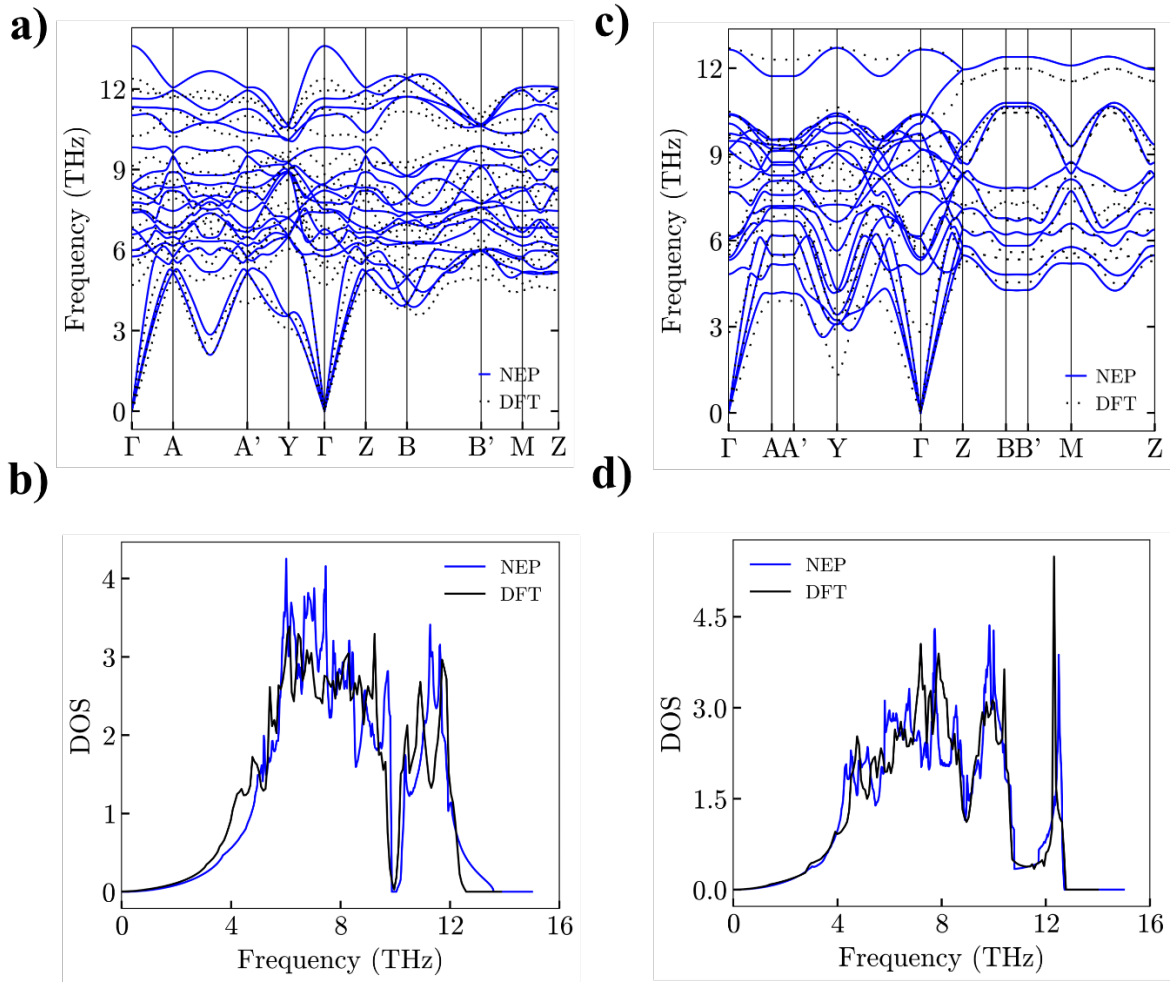

Figure 7. a) Phonon dispersion of C54 TiSi$_2$ ; b) Phonon dispersion of C49 TiSi$_2$; c) PDOS of C54 TiSi$_2$; and b) PDOS of C49 TiSi$_2$ from NEP and DFT.

### B. Thermal Transport in the Si/TiSi$_2$ Interface

The trained NEP is also capable of modeling the crystalline phases of TiSi$_2$ and its interfaces with Si. Crystalline silicide can be formed in devices when depositing Ti onto Si and performing high-temperature annealing. We examined NEP performance by performing NEMD simulations of Si/TiSi$_2$ interfaces for the commonly reported C54 and C49 phases. The calculated TBRs were $2.05 \pm 0.05$ and $1.93 \pm 0.02$ m$^2$K·GW$^{-1}$ for C49 and C54, respectively. We compared our C54 results with experiments conducted by Ye et al [15], who obtained conductances for the C54 phase on Si (100) and Si(111) surfaces as $2.2 \pm 0.05$ and $2.3 \pm 0.05$ m$^2$K·GW$^{-1}$, respectively, however the interfaces in that study were not flat, ideal interfaces as were modeled in the NEMD simulations. We observed that crystalline TiSi$_2$ interfaces with Si exhibit lower TBRs than sharp Si/Ti interfaces. The TBR is noticeably lower than the epitaxial Si/Ti interface, which we computed to be order of 2.3 m$^2$K·GW$^{-1}$ by NEMD simulations using NEP. The low values arise from the reduced phonon mismatch between crystalline TiSi$_2$ and Si, as evidenced by the PDOSs and spectral conductance plots in Fig. 8. Compared to Ti, SiTi$_2$ has

*Contact author: msingh96@gatech.edu

†Contact author: satish.kumar@me.gatech.edu

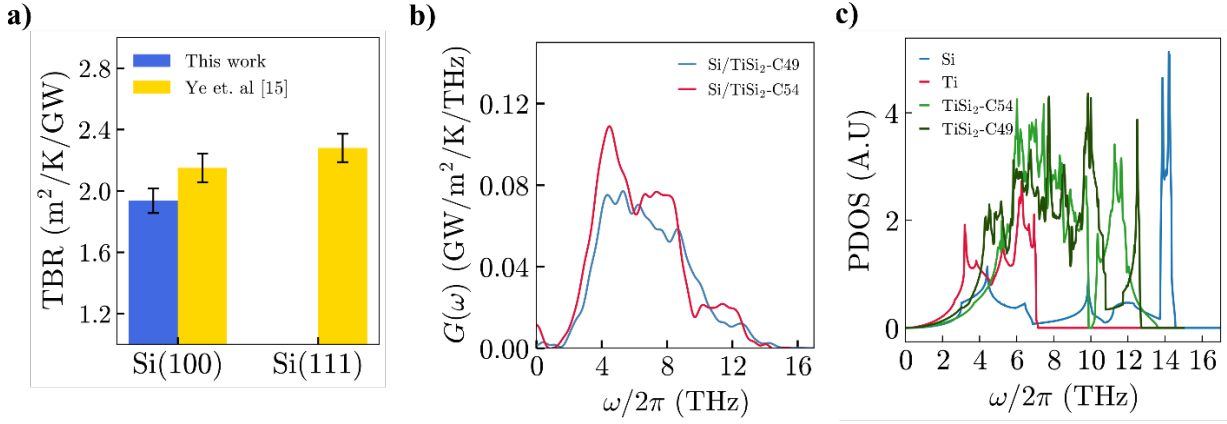

FIG. 8 a) Predicted TBRs at Si/TiSi$_2$ interfaces, compared with TBRs measured in [15] using TDTR; b) representative spectral conductance at the interface of Si/TiSi$_2$ for C54 and C49 TiSi$_2$; and c) PDOS of Ti, Si, and silicide; notice the larger range of PDOSs from metal silicide.

larger range of PDOS and good overlap with Si in the entire range of vibrational spectrum of Si up to 14 THz. The comparatively lower thermal resistance observed at TiSi$_2$ interfaces suggests that crystalline silicide layers could be an effective means to enhance thermal conductance in Si/Ti interfaces.

Further examination of the phonon spectral conductance reveals that within the 4 to 6 THz acoustic phonon range, the C54 interface exhibits notably higher phonon conductance compared to the C49 interface. This enhanced conductance in the lower-frequency acoustic modes (Fig. 8b) suggests improved vibrational coupling at the C54 interfaces, which is also evident in high peaks of PDOS in 4 to 6 THz range for C54 interface compared to the C49. Additionally, higher-frequency phonons, typically optical modes in the 6 to 10 THz range, also contribute significantly to thermal transport. The C54 interface shows elevated conductances relative to C49 except for a pronounced dip near 10 THz, which may correspond to the absence of PDOS in bulk C54 as shown in Fig. 7b. The phonon bandgap we observed in C54 TiSi$_2$ near 10 THz does not appear as zero spectral heat current, which suggests that anharmonic transport enables energy transmission close to this frequency.

### C. Thermal Transport: Effect of Silicide Layer at Ti/Si interface

We observed formation of a-TiSi$_x$ interlayer of considerable thickness at Si/Ti interface when these samples are fabricated by depositing Ti on Si and then annealed. To examine how this interfacial layer influences TBR, we performed simulations by sandwiching a-TiSi$_x$ between bulk Si and Ti and estimated the resulting TBR using NEMD. For the ideal and sharp Si/Ti [001] interface, we observed a low TBR of $2.31 \pm 0.05$ m$^2$K·GW$^{-1}$. The primarily contribution to thermal conductance for this sharp interface is from vibrational modes below 8 THz, which can be observed from the spectral conductance plot in Fig. 10 and is consistent with the strong overlap in PDOS of Si and Ti in this frequency range (Fig. 11a). We also observe significant contribution in the range of 10-14 THz where there is no PDOS in bulk Ti suggesting that inelastic scattering open channels for transmission of optical phonon modes in this range. We inserted an amorphous layer at the interface of varying thickness in the range of 0.5-2 nm and observed that as the amorphous interlayer thickness increases, the TBR also increases, reaching $4.94 \pm 0.07$ m$^2$K·GW$^{-1}$ at a thickness of ~2.0 nm, as shown in Fig. 9(c).

The addition of 0.5 nm amorphous silicide at the interface leads to overall increase in TBR to $2.7 \pm 0.04$ m$^2$KGW$^{-1}$ compared to $2.31 \pm 0.05$ m$^2$K·GW$^{-1}$ of a sharp Si/Ti interface. The addition of thin amorphous layer increases the thermal conductance contribution in the range of 8-10 THz (Fig. 10). Beyond this range, the phonon transport is significantly suppressed, especially in the frequency range of 3-8 THz, which leads to a significant increase in TBR. Even though there is a strong overlap of vibrational modes of a-TiSi with both Ti and Si (Fig. 11c), it could not provide a significant bridge for thermal transport, contrary to some prior observations [52], where a very thin amorphous layer is expected to enhance thermal conductance. The TBR increases linearly till the thickness of 1.5 nm of a-TiSi layer, but there is a sharp increase in TBR from 1.5 nm to 2 nm thickness (Fig. 9c), suggesting a major change in the transport mechanism. The spectral thermal conductance corresponding to 2 nm thick a-TiSi layer decreases for the entire range of frequency spectrum compared to sharp Si-Ti interface (Fig. 10). The primary transport carriers in amorphous material are diffusons, which can locally transfer the heat within the diffusion length. It is expected that when the thickness of a-TiSi layer is less than 1.5 nm, which is the order of the mean free path of diffusons, the a-TiSi still able to help in mediating thermal transport efficiently between Si and Ti, but beyond this length it acts more like a thermal barrier.

*Contact author: msingh96@gatech.edu

†Contact author: satish.kumar@me.gatech.edu

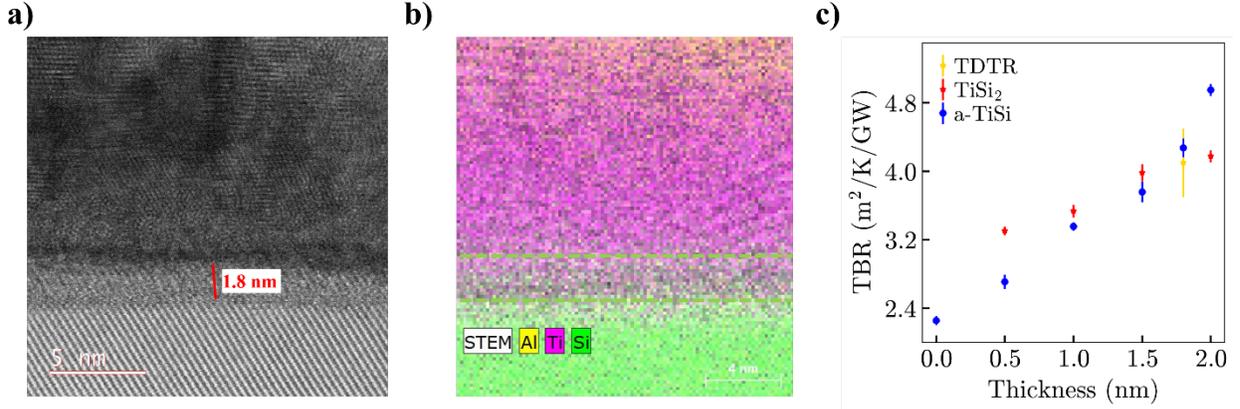

Figure 9. a) TEM image of an example Si/Ti sample with a thin amorphous layer at the interface between dashed green lines; b) EDS analysis of the interface; and c) thickness resolved TBR of Si/Ti interfaces, with amorphous or C54 silicide thin film inserts.

To validate the interfacial NEP model, simulation results for thin a-TiSi layer at the interface were compared against time-domain thermoreflectance (TDTR) measurements on Si/Ti/Al samples. The measured TBR values for this interface is $4.1 \pm 0.4 \, m^2 K \cdot GW^{-1}$, which is in good agreement with NEMD predictions of $4.26 \pm 0.1 \, m^2 K \cdot GW^{-1}$ for interfacial layer thicknesses near 1.8 nm. Rutherford backscattering spectrometry confirmed the composition of the metal layers, while transmission electron microscopy (TEM) and energy-dispersive X-ray spectroscopy (EDS), shown in Fig. 9a–b, revealed a $TiSi_x$ film thickness approaching 1.8 nm. This confirms the formation and spatial extent of the interfacial silicide layer and supports the conclusion that transport modeled by NEMD adequately captures the dominant thermal transport mechanisms at the Si/Ti interface. The high TBR obtained by the MLIP was obtained without using the electron-phonon correction proposed by Majumdar et al. [12]. This further indicates that when considering Si, in which phonons are primarily heat carrier, the effect of electron-phonon coupling does not have a significant effect on TBR at its interface with Ti. The approximate quantitative analysis of electron-phonon coupling can be performed, by expression:

$$G^{-1} = G_{pp}^{-1} + G_{ep}^{-1} \quad (6)$$

where $G_{pp}$ is the TBC obtained from NEMD simulations, and $G_{ep}$ is the conductance from electron-phonon coupling in the metal. $G_{ep}$ can be obtained from: $G_{ep} = \sqrt{g k_{ph}}$, where g is the electron-phonon conductance and $k_{ph}$ is the phonon thermal conductivity of the metal obtained from the first principles. Using this expression, we estimate less than 10% changes in TBR for Si/Ti interfaces.

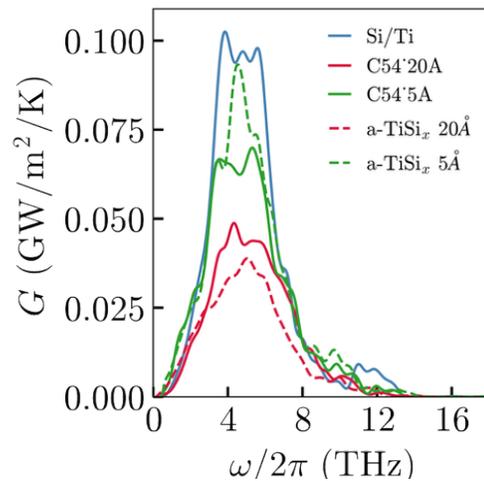

Figure 10. Spectral conductances of the Si/Ti with thin films of amorphous or C54 TiSi2.

*Contact author: msingh96@gatech.edu

†Contact author: satish.kumar@me.gatech.edu

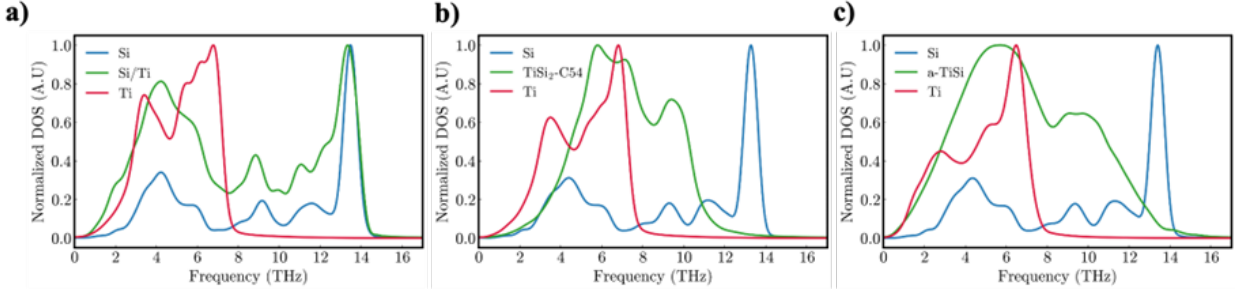

FIG 11. PDOS computed from NEMD simulations for (a) Si/Ti interface; (b) Si/C54/Ti; and (d) Si/a-TiSi/Ti. Si and Ti bulk PDOS are included for comparison.

To get further insights into the effect of a thin interfacial layer at Si/Ti interface, we estimated TBR and spectral conductance for different thicknesses of crystalline C54 TiSi$_2$ at the interface in the range of 0.5 to 2 nm and compared with a-TiSi$_x$ of similar thickness and observed some intriguing features. For 0.5 nm thickness of interfacial layer (IL), TBR corresponding to a-TiSi$_x$ is $2.7 \pm 0.08$ m$^2$K·GW$^{-1}$, which is lower than the TBR $3.3 \pm 0.05$ m$^2$KGW$^{-1}$ corresponding to crystalline TiSi$_2$ of the same thickness (Fig. 9c), and this trend continues till the thickness of the interfacial layer (IL) is 1.5 nm. At 2.0 nm thickness of the IL, the trend reverses, and the crystalline TiSi$_2$ IL has lower TBR of $4.17 \pm 0.1$ m$^2$K·GW$^{-1}$ compared to the a-TiSi$_x$ IL of $4.94 \pm 0.07$ m$^2$K·GW$^{-1}$. This indicates a very different transport mechanism at the interface when the thickness of IL is less than 2 nm. The spectral conductance shows a sharp peak in the range of 3-6 THz for the case of 0.5 nm a-TiSi$_x$ IL, which is absent for the crystalline IL of same thickness (Fig. 10). For the case of a-TiSi$_x$ IL, higher contributions to spectral conductance is also observed in the range of 8-10 THz (Fig. 10). Considering absence of PDOS in this range for Ti, it is clear indication that a-TiSi$_x$ helps in enabling anharmonic transport channels in this range but crystalline TiSi$_2$ does not. The broad vibrational spectrum of highly disordered a-TiSi$_x$ compared to crystalline SiTi$_2$ further supports this observation (Fig. 11c). When the thickness of IL increased to 2 nm, the spectral conductance for the case of a-TiSi$_x$ is less than for the crystalline SiTi$_2$ case for the entire frequency spectrum, suggesting that >2nm thicker amorphous layer could not act as efficient vibrational bridge for interfacial transport at the Si/Ti interface.

## V. CONCLUSIONS

A unified neuroevolution potential (NEP) for Si and Ti systems, extendable to modeling silicides and their interfaces, was developed from first principles. The NEP reproduces the phonon dispersions and density of states of both C49 and C54 TiSi$_2$ phases with good fidelity, capturing features such as phonon softening and band gaps. Using this framework, we performed NEMD simulations of Si/Ti interfaces along with the samples of crystalline and amorphous TiSi$_2$ interfacial layers of different thicknesses. The predicted TBR values show excellent agreement with TDTR experiments without considering electron-phonon coupling, directly challenging the assumption that electron–phonon coupling governs thermal transport across metal/semiconductor interfaces. The analysis establishes that a ~0.5 nm thin amorphous TiSi$_x$ structure at Si/Ti interface has reduced TBR compared to a crystalline TiSi$_2$ structure, but once the thickness of interfacial layer is higher than 1.5 nm, then the crystalline TiSi$_2$ interfacial layer leads to lower TBR, reversing the trend. We also compare TBRs at Si/TiSi$_2$ interfaces for different crystalline phases of TiSi$_2$ and observe that C54 phase has reduced TBR compared to C49 phase.

## ACKNOWLEDGMENTS

This work was supported by the Defense Advanced Research Projects Agency (DARPA) under the Thermal Modeling of Nanoscale Transistors (Thermonat) program.

*Contact author: msingh96@gatech.edu

†Contact author: satish.kumar@me.gatech.edu

[3] G.L. Pollack, Reviews of Modern Physics, Kapitza resistance, **41** (1), 48 (1969).

[4] W.A. Little, Canadian Journal of Physics, The transport of heat between dissimilar solids at low temperatures, **37** (3), 334 (1959).

[5] E.T. Swartz and R.O. Pohl, Thermal resistance at interfaces, Applied Physics Letters, **51** (26), 2200 (1987).

[6] J.C. Duda, J.L. Smoyer, P.M. Norris, and P.E. Hopkins, Extension of the diffuse mismatch model for thermal boundary conductance between isotropic and anisotropic materials, Applied Physics Letters, **95** (3), 031912 (2009).

[7] W. Zhang, T.S. Fisher, and N. Mingo, Simulation of interfacial phonon transport in Si-Ge heterostructures using an atomistic Green's function method, J Heat Trans-T Asme, **129** (4), 483 (2007).

[8] J. Dai and Z. Tian, Rigorous formalism of anharmonic atomistic Green's function for three-dimensional interfaces, Phys Rev B, **101** (4), 041301 (2020).

[9] L. Medrano Sandonas, R. Gutierrez, A. Pecchia, A. Croy, and G. Cuniberti, Quantum phonon transport in nanomaterials: Combining atomistic with non-equilibrium green's function techniques, Entropy, **21** (8), 735 (2019).

[10] S. Sadasivam, U.V. Waghmare, and T.S. Fisher, Phonon-eigenspectrum-based formulation of the atomistic Green's function method, Phys Rev B, **96** (17), 174302 (2017).

[11] Z.T. Tian, K. Esfarjani, and G. Chen, Enhancing phonon transmission across a Si/Ge interface by atomic roughness: First-principles study with the Green's function method, Phys Rev B, **86** (23), 235304 (2012).

[12] A. Majumdar and P. Reddy, Role of electron–phonon coupling in thermal conductance of metal–nonmetal interfaces, Applied Physics Letters, **84** (23), 4768 (2004).

[13] N. Mingo, Anharmonic phonon flow through molecular-sized junctions, Phys Rev B, **74** (12), 125402 (2006).

[14] S. Sadasivam, N. Ye, J.P. Feser, J. Charles, K. Miao, T. Kubis, and T.S. Fisher, Thermal transport across metal silicide-silicon interfaces: First-principles calculations and Green's function transport simulations, Phys Rev B, **95** (8), 085310 (2017).

[15] N. Ye, J.P. Feser, S. Sadasivam, T.S. Fisher, T. Wang, C. Ni, and A. Janotti, Thermal transport across metal silicide-silicon interfaces: An experimental comparison between epitaxial and nonepitaxial interfaces, Phys Rev B, **95** (8), 085430 (2017).

[16] A. Maiti, G.D. Mahan, and S.T. Pantelides, Solid State Commun, Dynamical simulations of nonequilibrium processes — Heat flow and the Kapitza resistance across grain boundaries, **102** (7), 517 (1997).

[17] T. Feng, Y. Zhong, J. Shi, and X. Ruan, Unexpected high inelastic phonon transport across solid-solid interface: Modal nonequilibrium molecular dynamics simulations and Landauer analysis, Phys Rev B, **99** (4), 045301 (2019).

[18] K. Khot, B. Xiao, Z. Han, Z. Guo, Z. Xiong, and X. Ruan, Phonon local non-equilibrium at Al/Si interface from machine learning molecular dynamics, Journal of Applied Physics, **137** (11), 115301 (2025).

[19] Q.S. Li, F. Liu, S. Hu, H.F. Song, S.S. Yang, H.L. Jiang, T. Wang, Y.K. Koh, C.Y. Zhao, F.Y. Kang, J.Q. Wu, X.K. Gu, B. Sun, and X.Q. Wang, Inelastic phonon transport across atomically sharp metal/semiconductor interfaces, Nat Commun, **13** (1), 4901 (2022).

[20] K.P. Wu, L. Zhang, D.B. Wang, F.Z. Li, P.Z. Zhang, L.W. Sang, M.Y. Liao, K. Tang, J.D. Ye, and S.L. Gu, A comparative study of interfacial thermal conductance between metal and semiconductor, Sci Rep-Uk, **12** (1), 19907 (2022).

[21] Y.X. Xu, H.Z. Fan, Z.G. Li, and Y.G. Zhou, Signatures of anharmonic phonon transport in ultrahigh thermal conductance across atomically sharp metal/semiconductor interface, International Journal of Heat and Mass Transfer, **201**, 123628 (2023).

[22] N. Yang, T.F. Luo, K. Esfarjani, A. Henry, Z.T. Tian, J. Shiomi, Y. Chalopin, B.W. Li, and G. Chen, Thermal interface conductance between aluminum and silicon by molecular dynamics simulations, J Comput Theor Nanos, **12** (2), 168 (2015).

[23] T.L. Feng, W.J. Yao, Z.Y. Wang, J.J. Shi, C. Li, B.Y. Cao, and X. Ruan, Spectral analysis of nonequilibrium molecular dynamics: Spectral phonon temperature and local nonequilibrium in thin films and across interfaces, Phys Rev B, **95** (19), 195202 (2017).

[24] K. Gordiz and A. Henry, Examining the effects of stiffness and mass difference on the thermal interface conductance between lennard-jones solids, Sci Rep-Uk, **5** (1), 18361 (2015).

[25] K. Sääskilahti, J. Oksanen, J. Tulkki, and S. Volz, Role of anharmonic phonon scattering in the spectrally decomposed thermal conductance at
*Contact author: msingh96@gatech.edu

†Contact author: satish.kumar@me.gatech.edu

*Contact author: msingh96@gatech.edu

†Contact author: satish.kumar@me.gatech.edu

*Contact author: msingh96@gatech.edu

†Contact author: satish.kumar@me.gatech.edu